\documentclass[aps,prd,twocolumn,showpacs,superscriptaddress,preprintnumbers,10pt]{revtex4-1}

\usepackage{amsmath,amsfonts,amssymb}
\usepackage{graphicx}
\usepackage{hyperref}
\usepackage{multirow}
\usepackage{nicefrac}
\usepackage{cases}
\usepackage{arydshln}

\usepackage[final]{showkeys}
\usepackage{mathrsfs}
\usepackage{stmaryrd}
\usepackage{dsfont}
\usepackage[Symbol]{upgreek}

\usepackage{tcolorbox}
\usepackage{empheq}
\usepackage{cancel}

\usepackage{color}
\definecolor{red}{rgb}{1,0,0}
\definecolor{lred}{rgb}{0.3,0,0}
\definecolor{green}{rgb}{0,0.6,0}
\definecolor{blue}{rgb}{0,0,1}
\definecolor{violet}{rgb}{0.8,0,0.8}

\definecolor{darkred}{rgb}{0.65,0.15,0}
\definecolor{darkgreen}{rgb}{.05,.5,.05}
\hypersetup{pdfborder={0 0 0},colorlinks=true,urlcolor=darkred,citecolor=darkred,linkcolor=darkred,linktocpage=true}

\usepackage{csquotes,textcase,xspace}

\usepackage{cases}
\usepackage{bm}
\usepackage{bbm}
\usepackage{multirow}
\usepackage{booktabs}
\usepackage{array}
\usepackage{tabularx}

\renewcommand{\d}{\ensuremath{\mathrm{d}}\xspace}

\newcommand{\SO}{\ensuremath{\mathrm{SO}}\xspace}

\newcommand{\vol}{{\,\rm vol}}
\def\sst#1{{\scriptscriptstyle #1}}

\def\0{{\sst{(0)}}}
\def\1{{\sst{(1)}}}
\def\2{{\sst{(2)}}}
\def\3{{\sst{(3)}}}
\def\4{{\sst{(4)}}}
\def\5{{\sst{(5)}}}
\def\6{{\sst{(6)}}}
\def\7{{\sst{(7)}}}
\def\8{{\sst{(8)}}}

\begin{document}

\preprint{MI-HET-811}
\title{On non-supersymmetric stable marginal deformations in AdS$_3$/CFT$_2$}
\author{Camille Eloy}
\email{camille.eloy@vub.be}
\affiliation{Theoretische Natuurkunde, Vrije Universiteit Brussel, and the International Solvay Institutes, Pleinlaan 2, B-1050 Brussels, Belgium}
\author{Gabriel Larios}
\email{gabriel.larios@tamu.edu}
\affiliation{Mitchell Institute for Fundamental Physics and Astronomy, Texas A\&M University, College Station, TX, 77843, USA}


\begin{titlepage}

\end{titlepage}

\begin{abstract}

We discuss a continuous family of non-supersymmetric AdS$_3\!\times\! S^3\!\times\! {\rm T^4}$ vacua in heterotic and type~II supergravities whose complete Kaluza-Klein spectrum is computed and found to be free from instabilities. This family is protected as well against some non-perturbative decay channels, and as such it provides the first candidate for a non-supersymmetric holographic conformal manifold in 2$d$. We also describe the operators realising the deformations in the worldsheet and boundary~CFT's.

\end{abstract}

\maketitle


Operators in any conformal field theory in $d$ dimensions can be classified according to their conformal dimension as relevant ($\Delta<d$), marginal ($\Delta=d$), or irrelevant ($\Delta>d$). While the cases with $\Delta\neq d$ trigger (possibly trivial) RG-flows between isolated CFTs, marginal operators play a different r\^ole: they describe the space of theories into which the original CFT can be deformed continuously without breaking conformal invariance.

In holographic CFTs, the bulk perspective over these conformal manifolds has remained an open challenge. The gravitational description of these deformations is given by families of AdS solutions that share the same cosmological constant and are labelled by free parameters. The main approach to building these solutions is provided by the TsT prescription \cite{Lunin:2005jy}, which can be applied whenever the undeformed solution preserves a number of abelian isometries (see \cite{Borsato:2018spz,Musaev:2023own} for novel approaches), and whenever the undeformed solution admits a consistent truncation down to a gauged supergravity in $d+1$ dimensions, the massless modes dual to the marginal operators usually sit at higher Kaluza-Klein levels \cite{Lunin:2005jy,Gauntlett:2005jb,Ahn:2005vc}, which prevents using the lower dimensional theory to describe the deformations -- see \cite{Cesaro:2021haf,Cesaro:2021tna,Giambrone:2021wsm} for a few recent exceptions to this rule.

In this work, we present a family of AdS${}_3$/CFT${}_2$ duals realising a two-dimensional conformal manifold labelled by parameters $(\omega,\zeta)$.
The undeformed solution is the AdS$_3\times S^3\times{\rm T}^4$ configuration of type II supergravities that preserves the small $\mathcal{N}=(4,4)$ superalgebra
\begin{equation}    \label{eq: 44superalgebra}
	\big[\text{SU}(2)_{\rm l}\ltimes \text{SU}(2\vert1,1)_{\rm L}\big]\times\big[\text{SU}(2)_{\rm r}\ltimes \text{SU}(2\vert1,1)_{\rm R}\big]\times{\rm U}(1)^{4}.
\end{equation}
This ten-dimensional solution only has a non-trivial profile for the fields in the NSNS sector, and can thus also be realised in heterotic string theory. For each of these ten-dimensional solutions, there are consistent truncations down to gauged supergravity in three dimensions, and unlike in \cite{Lunin:2005jy,Gauntlett:2005jb,Ahn:2005vc}, the scalar modes dual to the operators in the conformal manifold can already be captured within an appropriate truncation \cite{Eloy:2021fhc}.

Thanks to the reformulation of supergravity in the language of Exceptional Field Theory (ExFT), we construct the deformed solutions in $D=10$  by means of generalised Scherk-Schwarz (gSS) Ans\"atze. For generic values of the marginal deformations, the $10d$ spacetime is
\begin{equation}	\label{eq: topology2param}
	{\rm AdS}_3\times M^4\times{\rm T}^3\,,
\end{equation}
with the manifold $M^4$ (trivially) fibered over a deformed $S^3$ as $S_{y^7}^1\hookrightarrow M^4\rightarrow M^3_{\omega,\zeta}$,
with $y^7$ one of the coordinates on ${\rm T}^4$.
The generic deformations only preserve
\begin{equation}    \label{eq: symmgeneric}
     ({\rm U(1)}_{\rm L}\times{\rm U(1)}_{\rm R})\times{\rm SU(2)}_{\rm diag}\times{\rm U}(1)^4\,,
\end{equation}
and no supersymmetry, where ${\rm U(1)}_{\rm L,R}\subset{\rm SU(2)}_{\rm L,R}$ and ${\rm SU(2)}_{\rm diag}$ being the diagonal subgroup of ${\rm SU(2)}_{\rm l}\times{\rm SU(2)}_{\rm r}$ in \eqref{eq: 44superalgebra}.
Within the two-dimensional space of parameters, there is a line that preserves the $\mathcal{N}=(0,4)$ superalgebra
\begin{equation}    \label{eq: 04superalgebra}
	\text{U}(1)_{\rm L}\times\big[\text{SU}(2)_{\rm diag}\ltimes \text{SU}(2\vert1,1)_{\rm R}\big]\times{\rm U}(1)^{4}.
\end{equation}

Owing to the fact that ExFT's are not only a tool to describe consistent truncations, but encode the entire dynamics of the corresponding supergravities, we are also able to obtain the complete spectrum of Kaluza-Klein modes \cite{Malek:2019eaz} on the solutions we construct. This allows us to show that, despite generically non-supersymmetric, the two-parameter family of solutions is perturbatively stable for a finite range of the parameters.

In the remainder of this letter, we describe how these solutions appear in $D=3$ gauged supergravity and present their uplift both to heterotic and type II supergravities through the ExFT formalism. 
Subsequently, we discuss how the spectroscopy techniques of \cite{Malek:2019eaz,Eloy:2020uix} can be applied to these cases and the stability of the non-supersymmetric solutions.


The family of solutions found in \cite{Eloy:2021fhc} sits within the $3d$ half-maximal supergravity whose scalar manifold is 
\begin{equation}	\label{eq: so88coset}
	\frac{\rm SO(8,8)}{\rm SO(8)\times SO(8)}\,.
\end{equation}
The gauging procedure can be described by an embedding tensor $\Theta_{\bar K\bar L\vert \bar M\bar N}$, with the index $\bar M$ in the vector representation of $\SO(8,8)$. This embedding tensor must obey a quadratic constraint~\cite{Nicolai:2001ac,Deger:2019tem}
for the gauge algebra to close. Additionally, supersymmetry requires that the embedding tensor is restricted to take values~in~\cite{deWit:2003ja}
\begin{equation}
	\Theta\in\bm{1}\oplus\bm{135}\oplus\bm{1820}\,,
\end{equation}
and, therefore, it can be parameterised as
\begin{align}	\label{eq: so88theta}
	&\Theta_{\bar K\bar L\vert \bar M\bar N}=\\
	&\qquad\theta_{\bar K\bar L\bar M\bar N}+\frac12\big(\eta_{\bar M[\bar K}\theta_{\bar L]\bar N}-\eta_{\bar N[\bar K}\theta_{\bar L]\bar M}\big)+\theta\,\eta_{\bar N[\bar K}\eta_{\bar L]\bar M}\,,\nonumber
\end{align}
in terms of totally antisymmetric, symmetric traceless and singlet tensors, and with $\eta_{\bar M\bar N}$ the $\SO(8,8)$ invariant tensor. The embedding tensor describing our gauged supergravity is specified by the choice
\begin{equation}	\label{eq: so88Theta}
	\theta=0\,,
	\quad
	\theta_{\bar{0}\bar{0}\vphantom{\bar{\hat 2}}}=-4\sqrt{2}\,,
	\quad
	\theta_{\bar{M}\bar{N}\bar{P}\,\bar{0}}=-\tfrac1{\sqrt2}X_{\bar{M}\bar{N}\bar{P}}\,,
\end{equation}
with
\begin{equation}
		X_{\bar m\bar n\bar p}=
		X_{\bar m}{}^{\bar n\bar p}=
		X^{\bar m}{}_{\bar n}{}^{\bar p}=
		X^{\bar m\bar n}{}_{\bar p}=\varepsilon_{\bar m\bar n\bar p}	\,,
\end{equation}
in terms of indices following the breaking
\begin{equation}	\label{eq: so88intoso11gl3gl3so11}
	\begin{tabular}{ccc}
		$\SO(8,8)$	&	$\supset$	&	$\SO(1,1)\!\times\!{\rm GL(3)}\!\times\!{\rm GL(3)}\!\times\!\SO(1,1)$\,,	\\[5pt]
		$X^{\bar M}$		&	$\rightarrow$		&	$\{X^{\bar 0}, X_{\bar 0}, X^{\bar m}, X_{\bar m}, X^{\bar{\rm i}}, X_{\bar{\rm i}}, X^{\bar 7}, X_{\bar 7}\}$\,.
	\end{tabular}
\end{equation}
Here, the indices range as $\bar m\in\llbracket1,3\rrbracket$ and $\bar {\rm i}\in\llbracket4,6\rrbracket$, and for future convenience we also introduce $X^{\bar a}=\{X^{\bar{\rm i}},\ X^{\bar 7}\}$. A vacuum of this gauged supergravity is specified by a coset representative $\mathcal{V}$ in \eqref{eq: so88coset}, that extremises the scalar potential and defines the scalar matrix $M_{\bar K\bar L}=\mathcal{V}_{\bar K}{}^{\bar P}\mathcal{V}_{\bar L}{}^{\bar P}$.
In the basis \eqref{eq: so88intoso11gl3gl3so11} and with the $\mathfrak{so}(8,8)$ generators normalised as $(T^{\bar M\bar N}){}_{\bar P}{}^{\bar Q}=2\,\delta_{\bar P}{}^{[\bar M}\eta^{\bar N]\bar Q}$, the $(\omega,\zeta)$ solution can be characterised by
\begin{equation}	\label{eq: so88representative}
	\mathcal{V}_{\bar M}{}^{\bar N}=
		{\rm exp}\Big[-\omega\, T^{\bar 3}{}_{\bar 3}-\frac{\omega \zeta}{1-e^{-\omega}}\big(T^{\bar 3 \bar 7}-T^{\bar 3}{}_{\bar 7}\big)\Big]	\,,
\end{equation}
with all points sharing the AdS radius $\ell^2_{\rm AdS}=-2/{V_0}$.

To describe the ten-dimensional fields in a duality covariant language, we resort to $\SO(8,8)$ ExFT \cite{Hohm:2017wtr}, whose
bosonic fields are
\begin{equation}	\label{eq: SO88exftfields}
	\{g_{\mu\nu},\ \mathcal{M}_{MN},\ \mathcal{A}_{\mu}^{M N},\ \mathcal{B}_{\mu\, MN}\}\,,
\end{equation}
with $\mu\in\llbracket0,2\rrbracket$ and $M,N\in\llbracket1,16\rrbracket$ in the fundamental of $\SO(8,8)$. All these fields depend on both external coordinates $x^{\mu}$ and internal ones $Y^{MN}$, with the latter in the adjoint of $\SO(8,8)$. The 7-dimensional internal coordinates $y^{i}$ parametrising the three-sphere and torus are embedded in $Y^{MN}$. To ensure that the fields depend only on $y^{i}$, the coordinate dependance is subject to the section constraints
\begin{equation} \label{eq:sectionSO88}
	\partial_{[MN}\otimes\partial_{PQ]}=0, \quad \eta^{PQ}\partial_{MP}\otimes\partial_{NQ}=0,
\end{equation}
which can be solved by breaking
\begin{equation}	\label{eq: so88pBreaking}
	\begin{tabular}{ccc}
		$\SO(8,8)$	&	$\supset$	&	$\SO(1,1)\times{\rm GL}(7)$\,,	\\[5pt]
		$X^M$		&	$\longrightarrow$		&	$\{\ X^0,\ X_0,X^i,\ X_i\}$\,,
	\end{tabular}
\end{equation}
and restricting coordinate dependance to $y^i=Y^{i0}$. We align ExFT indices with the ones in the three-dimensional theory by embedding ${\rm GL}(3)\times{\rm GL}(3)\times{\rm SO}(1,1)\subset {\rm GL}(7)$ as in \eqref{eq: so88intoso11gl3gl3so11}.

The explicit dictionary between the $\SO(8,8)$-ExFT generalised metric and the internal components of the NSNS fields is given by
\begin{gather}
	 \mathcal{M}^{00}=\hat{g}^{-1}e^{\hat\Phi/2}\,,	\quad
	\mathcal{M}^{0i}=\tfrac{1}{6!}\,\mathcal{M}^{00}\varepsilon^{ij_1\dots j_6}\, \tilde{b}_{j_1\dots j_6}\,,			\nonumber\\[4pt]
	\mathcal{M}^{00}\mathcal{M}^{ij}-\mathcal{M}^{0i}\mathcal{M}^{0j}=\hat{g}^{-1}\hat{g}^{ij}\,,				\label{eq: so88ExFTdictionary} \\[4pt]
	\mathcal{M}^{00}\mathcal{M}^{i}{}_{j}-\mathcal{M}^{0i}\mathcal{M}^{0}{}_{j}=\hat{g}^{-1}\hat{g}^{ik}\, b_{kj}\,,	\nonumber
\end{gather}
where $\hat{g}_{ij}$ is the purely internal block of the ten-dimensional metric in Einstein frame, and $\hat{g}$ its determinant. The fields $b$ and $\tilde{b}$ are not directly related to the higher-dimensional two-form, but retrieve its field strength as
\begin{equation}
	\hat H=\d b+e^{\hat\Phi/8}\star_{10}\d\tilde{b}\,.
\end{equation}

Upon solving the section conditions, contact with gauged supergravity is  achieved through the gSS Ansatz
\begin{align}	\label{eq: so88SSansatz}
	g_{\mu\nu}(x,Y)&=\rho^{-2}g_{\mu\nu}(x)\,,	\nonumber\\[4pt]
	\mathcal{M}_{MN}(x,Y)&=U_{M}{}^{\bar M}U_{N}{}^{\bar N}M_{\bar M\bar N}(x)\,,	\\[4pt]
	\mathcal{A}_\mu^{MN}(x,Y)&=\sqrt{2}\,\rho^{-1}(U^{-1})_{\bar M}{}^{M}(U^{-1})_{\bar N}{}^{N}A_\mu^{\bar M\bar N}(x)\,,			\nonumber\\[4pt]
	\mathcal{B}_{\mu KL}(x,Y)&=-\frac{\rho^{-1}}{2\sqrt{2}}\,U_{M \bar N}\partial_{KL}(U^{-1})_{\bar M}{}^{M}A_\mu^{\bar M\bar N}(x)\,,		\nonumber
\end{align}
with $\rho(Y)$ a scale factor and $U_{M}{}^{\bar M}(Y)$ an element of $\SO(8,8)$ controlling the twisting of the 3$d$ metric, vectors and scalars by the internal coordinates. The relevant pair $(\rho,\, U)$ which recovers \eqref{eq: so88Theta} can be constructed out of the SO(4,4)-ExFT parallelisation discussed in \cite{Eloy:2021fhc} by embedding it in the $\{X^0,X_0,X^m,X_m\}$ block in \eqref{eq: so88intoso11gl3gl3so11}.

The solutions advertised in~\eqref{eq: topology2param} then follow from introducing the Ansatz~\eqref{eq: so88SSansatz} with the 3$d$ representative~\eqref{eq: so88representative} in~\eqref{eq: so88ExFTdictionary}.
We choose our coordinates as
\begin{equation}
	\begin{aligned}
		&Y^{m,0}=\{\cos\alpha\,\cos\beta\,,	\cos\alpha\,\sin\beta\,, \sin\alpha\,\cos\gamma\}\,,	\\
		&Y^{a,0}=y^a\,,
	\end{aligned}
\end{equation}
with $y^a\sim y^a+1$ parameterising ${\rm T}^{4}$, and the angles $0\leq\alpha\leq\frac\pi2$ and $0\leq\beta,\gamma\leq 2\pi$ 
describing a deformed three-sphere with metric
\begin{align}
	&\d s^2(M^3_{\omega,\zeta})=																			\nonumber\\[5pt]
	&\qquad\d\alpha^2+ e^{\omega}\Delta^4\big(\cos^2\!\alpha\,\d\beta^2+(\zeta^2+e^{-2\omega})\sin^2\!\alpha\,\d\gamma^2\big) \nonumber\\[5pt]
	&\qquad\quad-e^{2\omega}\zeta^2\Delta^8\,\big(\cos^2\!\alpha\,\d\beta-\sin^2\!\alpha\,\d\gamma\big)^2\,.					
\end{align}
In terms of these coordinates, the solution reads
\begin{align}	\label{eq: twoparamin10D}
	e^{\hat\Phi}&=\Delta^2,	\nonumber\\[5pt]
	\d\hat{s}^2_{\rm s}&= \d s^2({\rm AdS}_3)+\d s^2(M^3_{\omega,\zeta})+\delta_{\rm ij}\,\d y^{\rm i}\d y^{\rm j}		\nonumber\\[5pt]
	&\quad+\big[\d y^{7}+e^{\omega}\zeta\Delta^4\,\big(\cos^2\!\alpha\,\d\beta-\sin^2\!\alpha\,\d\gamma\big)\big]^2,		\nonumber\\[5pt]
	\hat{H}_{(3)}&= 2\vol({\rm AdS}_3)+\sin(2\alpha)\,\Delta^8e^{2\omega}	\\
	&\quad\times\d\alpha\wedge(\d\beta+\zeta\d y^7)\wedge\big((\zeta^2+e^{-2\omega})\d\gamma-\zeta\d y^7\big)\,, 	\nonumber
\end{align}
with the function
\begin{equation}	\label{eq: warping6D2param}
	\Delta^2=\frac{e^{-\omega/2}}{\sqrt{1+(\zeta^2+e^{-2\omega}-1)\cos^2\!\alpha}}\,,
\end{equation}
and the string frame metric in \eqref{eq: twoparamin10D} given by $\hat{g}_{\rm s}{}_{\hat\mu\hat\nu}\!=~\!\!\!e^{\hat\Phi/2}\hat{g}_{\hat\mu\hat\nu}$.
For generic values of the marginal parameters, the solution preserves \eqref{eq: symmgeneric}, with the abelian factors acting as shifts on $\beta$, $\gamma$ and the angles on the torus, and ${\rm SU}(2)_{\rm diag}$ as rigid rotations preserving $\delta_{\rm ij}\d y^{\rm i}\d y^{\rm j}$.

Effectively, the $\zeta$ modulus controls the fibration of $M^4$ in~\eqref{eq: topology2param}.  
When setting $\zeta^2=1-e^{-2\omega}$, 
$M^3_{\omega,\zeta}$ itself becomes a Hopf fibration, $S_{\theta}^1\hookrightarrow M^3_{\omega}\rightarrow \mathbb{CP}^1$, and
the family of solutions in \eqref{eq: twoparamin10D} simplifies to
\begin{align}	\label{eq: susysolution}
	\hat\Phi&=-\frac\omega2,		\nonumber\\[5pt]
	\d\hat{s}_{\rm s}^2&= \d s^2\big({\rm AdS}_3\big) + \delta_{\rm ij}\,\d y^{\rm i}\d y^{\rm j}+\d s^2\big(\mathbb{CP}^1\big)+e^{-2\omega}\,\bm{\eta}^2	\nonumber\\
	&\quad+\big(\d y^{7}+\sqrt{1-e^{-2\omega}}\,\bm{\eta}\big)^2\,,	\\[5pt]
	\hat{H}_{(3)}&= 2\vol({\rm AdS}_3)+2\,\bm{\eta}\wedge\bm{J}+2\sqrt{1-e^{-2\omega}}\,\bm{J}\wedge \d y^7\,,	\nonumber
\end{align}
which, away from the scalar origin, preserves the $\mathcal{N}\!=~\!\!\!(0,4)$ superalgebra in \eqref{eq: 04superalgebra}. The ${\rm SU}(2)_{\rm R}$ there is realised as the isometries of the Fubini-Study metric on $\mathbb{CP}^1$, and ${\rm U(1)}_{\rm L}$ as shifts of the angle $\theta$ along the Hopf fibre. Here, we define 
\begin{equation}
	\bm{\eta}=\cos^2\!\alpha\,\d\beta-\sin^2\!\alpha\,\d\gamma\,, 
\end{equation}
together with $\bm{J}$ and $\bm{\Omega}$, who are respectively the contact, K\"ahler and complex holomorohic forms of the Sasaki-Einstein structure on $S^3$. They satisfy
\begin{equation} \label{eq: SE3}
	\begin{gathered}
		\d\bm{\eta}=2\bm{J}\,,	\qquad\quad \d\bm{\Omega}=2i\bm{\eta}\wedge\bm{\Omega}\,,\\
        \bm{J}\wedge\bm{\Omega}=0\,,		\qquad    \bm{\eta}\wedge\bm{J}=\frac{i}2\bm{\eta}\wedge\bm{\Omega}\wedge\bm{\bar\Omega}=\vol(S^3)\,.
    \end{gathered}
\end{equation}
The solution~\eqref{eq: susysolution} is analogous to the ${\cal N}=4$ vacua found in~\cite{Eloy:2023zzh} in the context of AdS$_3\times S^3\times S^3\times S^1$.
Recently, similar solutions have appeared in \cite{Lozano:2019emq,Lima:2022hji,Balaguer:2021wjf}. However, those solutions require the presence of D-branes which sit outside the S-duality orbit of our purely F1-NS5 configuration. The presence of the aforementioned fibrations has also been an obstacle for obtaining them as the near horizon limit of brane intersection with flat branes.

From a string worldsheet perspective, the configuration \eqref{eq: twoparamin10D} can be described as a deformation of the ${\rm SL}(2,\mathbb{R})\times{\rm SU}(2)\times{\rm U}(1)^4$ WZW model \cite{Maldacena:2000hw,Giveon:1998ns}
corresponding to the undeformed solution \footnote{We would like to thank E. Martinec for this helpful observation}. Focusing on \eqref{eq: susysolution} for simplicity, the operator controlling the deformation is $J^z_{{\rm SU}(2)}\bar{J}_{{\rm U}(1)_7}$, where $J^z_{{\rm SU}(2)}$ is a component of the holomorphic current realising the left-moving copy of ${\rm SU}(2)$ in the symmetry group, and $\bar{J}_{{\rm U}(1)_7}$ the anti-holomorphic current corresponding to shifts in $y^7$. Being the product of conserved (anti-)holomorphic commuting currents, the operator $J^z_{{\rm SU}(2)}\bar{J}_{{\rm U}(1)_7}$ is exactly marginal~\cite{Chaudhuri:1988qb} and breaks the superalgebra from \eqref{eq: 44superalgebra} to \eqref{eq: 04superalgebra}. Analogously, in the conformal ${\rm Sym}^N({\rm T}^4)$ theory conjectured to sit at the boundary of ${\rm AdS}_3$ \cite{Maldacena:1997re,Eberhardt:2018ouy}, the single-particle operator realising the deformation can be identified~as 
\begin{equation}	\label{eq: opT4cft}
	\mathcal{O}\sim\sum_k^N (j^z_{{\rm SU}(2)}\bar{j}_{S^1_7})_k\,,
\end{equation}
with now $j^z_{{\rm SU}(2)}$ a component of the left-moving R-symmetry group and $\bar{j}_{S^1_7}$ one of the right-moving currents of ${\rm T}^4$. The sum in \eqref{eq: opT4cft} assures that this operator survives the orbifold projection. Similar considerations can be made for $\eqref{eq: twoparamin10D}$, with the deformations now breaking supersymmetry completely.


\vspace{10pt}

If one perturbs the Scherk-Schwarz parallelisation in~\eqref{eq: so88SSansatz} \`a la \cite{Malek:2019eaz}, the spectrum of modes that only excite NSNS fields can be retrieved. We thus consider the following expansion in~\eqref{eq: so88SSansatz}~\cite{Eloy:2020uix}:
\begin{equation}	\label{eq: so88SSansatzPert}
	\begin{aligned}
	g_{\mu\nu}(x)&\rightarrow \bar{g}_{\mu\nu}(x)+h_{\mu\nu}{}^{\Lambda(p_a)}(x)\mathcal{Y}^{\Lambda(p_a)}\,,	\\[4pt]
	M_{\bar M\bar N}(x)&\rightarrow \bar{M}_{\bar M\bar N}+j_{\bar M\bar N}{}^{\Lambda(p_a)}(x)\mathcal{Y}^{\Lambda(p_a)}\,, \\[4pt]
	A_\mu^{\bar M\bar N}(x)&\rightarrow a_\mu^{\bar M\bar N\, \Lambda(p_a)}(x)\mathcal{Y}^{\Lambda(p_a)}\,,
	\end{aligned}
\end{equation}
where the background is described by
\begin{equation}	\label{eq: so88exftbkg}
	\{g_{\mu\nu},\ M_{\bar M\bar N},\, A_\mu^{\bar M\bar N}\}=\{\bar{g}_{\mu\nu},\ \bar{M}_{\bar M\bar N},\, 0\}\,,
\end{equation}
and $\{h_{\mu\nu}{}^{\Lambda(p_a)},\ j_{\bar M\bar N}{}^{\Lambda(p_a)},$ $a_\mu^{\bar M\bar N\, \Lambda(p_a)}\}$ are  the perturbations expanded in a basis of scalar harmonics of the $S^3\times{\rm T}^4$ configuration that preserves maximal isometry. As such, they furnish the infinite-dimensional reducible representation of $\SO(4)\times{\rm U(1)}^4$
\begin{equation}	\label{eq: Yfactorisation}
	\mathcal{Y}^{\Lambda(p_a)}=\mathcal{Y}^{\Lambda}\,e^{2\pi i\sum p_ay_a}
	\in\bigoplus_{p_a \in \mathbb{Z}^4}\;\bigoplus_{n=0}^{\infty}\Big(\frac n2,\frac n2\Big)_{(p_a)}\,.
\end{equation}
Here, $\Lambda$ denotes Kaluza-Klein index on $S^3$, which can be expanded as
\begin{equation}	\label{eq: tower_harm}
	\mathcal{Y}^{\Lambda}=\big\{1,\ \mathcal{Y}^{\alpha},\ \mathcal{Y}^{\{\alpha}\mathcal{Y}^{\beta\}},\,\dots\big\}\,,
	\qquad \alpha\in\llbracket1,4\rrbracket\,,
\end{equation}
with braces denoting traceless symmetrization. This leads to the definition of $\mathring{\mathcal{T}}_{\bar M\bar N}{}^{(p_a)\Lambda\, \Sigma}$ as the representation matrix encoded in the $\SO(8,8)$ twist matrix~as
\begin{equation}
	\begin{aligned}
		\rho^{-1}(U^{-1})_{\bar M}{}^{M}(U^{-1})_{\bar N}{}^{N}&\partial_{MN}\mathcal{Y}^{\Lambda(p_a)}=\\
		&-\sqrt2\,\mathring{\mathcal{T}}_{\bar M\bar N}{}^{(p_a)\Lambda\Sigma}\mathcal{Y}^{\Sigma(p_a)}\,,
	\end{aligned}
\end{equation}
which, following \eqref{eq: Yfactorisation}, can in turn be decomposed as
\begin{equation}
	\mathring{\mathcal{T}}_{\bar M\bar N}{}^{(p_a)\Lambda\Sigma}=\mathring{\mathcal{T}}_{\bar M\bar N}{}^{\Lambda\Sigma}+\delta^{\Lambda\Sigma}\,\mathring{\mathcal{T}}_{\bar M\bar N}{}^{(p_a)}\,.
\end{equation}
For our twist, the SO(4) piece $\mathring{\mathcal{T}}_{\bar M\bar N}{}^{\Lambda\Sigma}$ has non-vanishing components
\begin{equation}	\label{eq: curlyTkk1SO88S3}
	\mathring{\mathcal{T}}\,_{\bar m\bar0}\,{}^{\alpha\beta}=\sqrt{2}\,\delta^{[\alpha}_{4}\delta^{\beta]}_{\bar m}\,,		\qquad
	\mathring{\mathcal{T}}\,^{\bar m}{}_{\bar0}\,{}^{\alpha\beta}=\dfrac{1}{\sqrt{2}}\,\varepsilon^{\bar{m}4\alpha\beta}\,,
\end{equation}
when acting on the level $n=1$ harmonics. 
Higher level tensors can then be constructed recursively from \eqref{eq: curlyTkk1SO88S3} \cite{Eloy:2021fhc}. %
Similarly, the ${\rm U}(1)^4$ block is given by
\begin{equation}	\label{eq: curlyTso2}
	\mathring{\mathcal{T}}\,_{\bar a\bar0}\,{}^{(p_a)}=-\frac1{\sqrt2}\,2\pi i\, p_a\,.
\end{equation}

Introducing \eqref{eq: so88SSansatzPert} into the ExFT equations of motion and keeping only terms linear in the perturbations, one can read off mass matrices whose eigenvalues, modulo removal of redundancies and Goldstone modes~\cite{Eloy:2021fhc}, are the masses of the modes in the KK spectrum that only excite NSNS fields.
To additionnaly capture perturbations exciting vectors in the heterotic theory, we can embed $\SO(8,8)$ into $\SO(8,24)$ with trivial components on the $\SO(16)$ block. On the other hand, to describe the modes that excite RR fields of type~II supergravities, the $\SO(8,8)$ theory must be embedded in E$_{8(8)}$ as
\begin{equation}	\label{eq: E8intoSO88}
\begin{tabular}{ccc}
	${\rm E}_{8(8)}$	&$\supset$&	SO(8,8)\,,	\\[5pt]
	$\bm{248}$		&$\to$&		$\bm{120}+\bm{128}_{\rm s}$\,,		\\[5pt]
	$t^{{\mathcal{M}}}$	&$\to$&		$\{t^{[ M N]},\ t^{{\mathcal{A}}}\}$\,,
\end{tabular}
\end{equation}
and analogously for barred indices. The relevant $3d$ gauged supergravity is described by a symmetric embedding tensor $X_{\cal \bar{M}\bar{N}}$ living in the $\bm{1}\oplus\bm{3875}$ representation of E$_{8(8)}$ and subject to \cite{Nicolai:2003bp,Hohm:2005ui}
\begin{equation}	\label{eq: E88quadconstr}
	X_{\bar{\mathcal{R}}\bar{\mathcal{P}}}\,X_{\bar{\mathcal{S}}(\bar{\mathcal{M}}}\,f_{\bar{\mathcal{N}})}{}^{\bar{\mathcal{R}}\bar{\mathcal{S}}}=0\,,
\end{equation}
with $f_{\bar{\mathcal{N}}\bar{\mathcal{R}}}{}^{\bar{\mathcal{S}}}$ the E$_{8(8)}$ structure constants. Taking the latter as
\begin{equation}	 \label{eq: e88fs}
	\begin{aligned}
		f_{MN,PQ}{}^{RS} &= -8\,\delta_{[M}{}^{[R}\eta_{N][P}\delta_{Q]}{}^{S]}\,,\\
		f_{MN,\mathcal A}{}^{\mathcal B} &= \frac{1}{2}\,\Gamma_{MN\,\mathcal A}{}^{\mathcal B}\,,\\
		f_{\mathcal A\mathcal B}{}^{MN}&=-\frac{1}{2}\,\Gamma^{MN}_{\mathcal A\mathcal B}\,,
	\end{aligned}
\end{equation}
the ${\rm E}_{8(8)}$ quadractic constraint~\eqref{eq: E88quadconstr} is solved by embedding the $\SO(8,8)$ embedding tensor~\eqref{eq: so88Theta} in $X_{\cal \bar{M}\bar{N}}$ as~\cite{Deger:2019tem}
\begin{equation}	\label{eq: EmbTensorDecompositionE8}
	\begin{aligned}
		X_{\bar K\bar L\vert \bar M\bar N}&=2\,\Theta_{\bar K\bar L\vert \bar M\bar N}\,,	\\
		X_{\bar{\mathcal{A}}\bar{\mathcal{B}}}&=-\theta\,\eta_{\bar{\mathcal{A}}\bar{\mathcal{B}}}+\tfrac1{48}\,\Gamma_{\bar{\mathcal{A}}\bar{\mathcal{B}}}^{\bar K\bar L\bar M\bar N}\theta_{\bar K\bar L\bar M\bar N}\,,
	\end{aligned}
\end{equation}
in terms of the chiral gamma matrices of $\SO(8,8)$ and the charge conjugation matrix $\eta_{\bar{\mathcal{A}}\bar{\mathcal{B}}}$. The $(\omega,\zeta)$ family of solutions is then characterised by the ${\rm E}_{8(8)}/\SO(16)$ representative
\begin{equation}	\label{eq: e88representative}
		\mathcal{V}_{\cal\bar M}{}^{\cal\bar N}=
		{\rm exp}\Big[-\omega\, f^{\bar 3}{}_{\bar 3}-\frac{\omega \zeta}{1-e^{-\omega}}\big(f^{\bar 3 \bar 7}-f^{\bar 3}{}_{\bar 7}\big)\Big]\,.
\end{equation}

This maximal gauged supergravity can be uplifted to $10d$ by means of E$_{8(8)}$ ExFT \cite{Hohm:2014fxa}, whose fields are
\begin{equation}	\label{eq: e88exftFields}
	\{g_{\mu\nu},\ \mathcal{M}_{\mathcal{M}\mathcal{N}},\ \mathcal{A}^{\mathcal{M}}_{\mu},\ \mathcal{B}_{\mu\, \mathcal{M}}\}\,,
\end{equation}
alongside their fermionic superpartners \cite{Baguet:2016jph}. All these fields depend on both the same three external coordinates $x^{\mu}$ as before, as well as on a set of 248 extended coordinates $Y^{\mathcal{M}}$. Coordinate dependence is however restricted by the section constraints
\begin{align}	\label{eq:sectionE8}
	\kappa^{\mathcal{M}\mathcal{N}}\partial_{\mathcal{M}}\otimes\partial_{\mathcal{N}} &= 0\,,				\nonumber\\
	f^{\mathcal{M}\mathcal{N}}{}_{\mathcal{P}}\partial_{\mathcal{M}}\otimes\partial_{\mathcal{N}} &= 0\,,		\\
	(\mathbb{P}_{3875})_{\mathcal{M}\mathcal{N}}{}^{\mathcal{K}\mathcal{L}}\partial_{\mathcal{K}}\otimes\partial_{\mathcal{L}} &= 0\,,\nonumber
\end{align}
acting, as usual, on any combination of fields or gauge parameters. The Cartan-Killing form $\kappa_{\cal MN}$ and projector $(\mathbb{P}_{3875})_{\mathcal{M}\mathcal{N}}{}^{\mathcal{K}\mathcal{L}}$ can be found in \cite{Hohm:2005ui}. The uplift can be expressed as the gSS factorisation
\begin{align}
	g_{\mu\nu}(x,Y)&=\rho^{-2}g_{\mu\nu}(x)\,, \nonumber\\[4pt]
	\mathcal{M}_{\cal MN}(x,Y)&=U_{\cal M}{}^{\cal \bar M}U_{\cal N}{}^{\cal \bar N}M_{\cal \bar M\bar N}(x)\,, \nonumber\\[4pt]
	\mathcal{A}_{\mu}{}^{\cal M}(x,Y)&=\rho^{-1}(U^{-1})_{\cal \bar M}{}^{\cal M}\,A_\mu^{\cal \bar M}(x)\,,  \label{eq: e88SSansatz}		\\[4pt]
	\mathcal{B}_{\mu {\cal M}}(x,Y)&=\frac{\rho^{-1}}{60}\,f_{\cal \bar M}{}^{\cal \bar P\bar Q}\,(U^{-1})_{\cal \bar PP}\partial_{\cal M}(U^{-1})_{\cal \bar Q}{}^{\cal P}\,A_\mu^{\cal \bar M}(x)\,.  \nonumber
\end{align}

As we did for the embedding tensor in~\eqref{eq: EmbTensorDecompositionE8}, we solve the section constraints and parametrise the twist matrix by using the $\SO(8,8)$ setup detailed above. We embed the coordinates as $y^{i}\subset Y^{MN}\subset Y^{\cal M}$ and discard all dependencies on $Y^{\cal A}$, so that the ${\rm E}_{8(8)}$ section conditions~\eqref{eq:sectionE8} follow from~\eqref{eq:sectionSO88}. Concerning the uplift, we use the same $\rho$ as before, and the twist matrix
\begin{equation}	\label{eq: twistE88}
	U_{\mathcal{M}}{}^{\bar{\mathcal{M}}}=
	\left(
		\begin{array}{cc}
			U_{[M}{}^{\bar{M}}U_{N]}{}^{\bar{N}}		&	0							\\
			0							&	U_{\mathcal{A}}{}^{\bar{\mathcal{A}}}	\\
		\end{array}
	\right).
\end{equation}
$U_{\mathcal{A}}{}^{\bar{\mathcal{A}}}$ is a $\bm{128}_{\rm s}$ representation of $U_{M}{}^{\bar{M}}\in\SO(8,8)$:
\begin{equation}
	U_{\mathcal{A}}{}^{\bar{\mathcal{A}}}=\exp\left(\frac{1}{2}\,u_{MN}\,\Gamma^{MN}\right)_{\mathcal{A}}{}^{\bar{\mathcal{A}}},
\end{equation}
where $u$ is such that $U_{M}{}^{\bar M}=\exp\left(u_{PQ}\,T^{PQ}\right)_{M}{}^{\bar M}$. 

The KK modes can then be captured by perturbing the Ansatz \eqref{eq: e88SSansatz} as
\begin{equation}	\label{eq: e88SSansatzfluct}
	\begin{aligned}
	g_{\mu\nu}(x)&\to {\bar g}_{\mu\nu}(x)+h_{\mu\nu}{}^{\Lambda(p_a)}(x)\,{\cal Y}^{\Lambda(p_a)}\,, \\[4pt]
	M_{\cal \bar M\bar N}(x)&\to \bar{M}_{\cal \bar M\bar N}+j_{\cal \bar M\bar N}{}^{\Lambda(p_a)}(x)\,{\cal Y}^{\Lambda(p_a)}\,,	\\[4pt]
	A_\mu^{\bar {\cal M}}(x)&\to a_\mu^{\bar {\cal M},\,\Lambda(p_a)}(x)\,{\cal Y}^{\Lambda(p_a)}\,,
	\end{aligned}
\end{equation}
with the harmonics in \eqref{eq: Yfactorisation}.
Again, thanks to the choice of harmonics, the mass operators that can be read off from the linearised equations of motion in ExFT become algebraic matrices, given that
\begin{equation}
	\rho^{-1}(U^{-1})_{\cal \bar M}{}^{\cal M}\partial_{\cal M}{\cal Y}^{\Lambda(p_a)} = -\,{\cal T}_{\cal\bar M}{}^{(p_a)\Lambda\Sigma}\,{\cal Y}^{\Sigma(p_a)}\,,
\end{equation}
with only non-vanishing components ${\cal T}_{\bar M\bar N} = 2\,\mathring{\cal T}_{\bar M\bar N}$.
Further details on these E$_{8(8)}$-covariant Kaluza-Klein mass matrices will be given elsewhere~\cite{EloyGalliItsiosLariosMalek}.


Armed with the ExFT mass matrices for the KK modes, we have computed the masses in the different 3$d$ supergravities and ExFT's for the first few levels on the $S^3$ and arbitrary level on the T$^4$ for bosons and fermions. These results can be encapsulated in a simple master formula in terms of the charges of the modes under the relevant symmetry (super-)algebra.

The Kaluza-Klein spectrum of type II supergravities on the round AdS$_3\times S^3\times{\rm T}^4$ organises into supermultiplets of~\eqref{eq: 44superalgebra}. We denote by $p_{a}$ the ${\rm U}(1)^4$ charges, and long multiplets of $\text{SU}(2)\ltimes \text{SU}(2\vert1,1)$ as $\left[\Delta,j^{-},j^{+}\right]$ -- see the appendix~A of ref.~\cite{Eloy:2021fhc} for a review -- with $\Delta$ the conformal dimension of the conformal primary and $j^{+}, j^{-}$ its spins under the two ${\rm SU}(2)$ factors. The type II spectrum at this point is given by (\emph{c.f.}~\cite{deBoer:1998kjm})
\begin{equation} \label{eq:spectrumSU211}
 	\mathcal{S}=\bigoplus_{\substack{j^{+}\geq0\\p_{a}\in\mathbb{Z}^4}}\Big(\big[\Delta_{\rm L},0,j^{+}\big]\otimes\big[\Delta_{\rm R},0,j^{+}\big]\Big)_{\{p_{a}\}}\,,
\end{equation}
where the conformal dimension of the primary of each factor is
\begin{equation} \label{eq:confdimorigin}
	\Delta_{\rm L}=\Delta_{\rm R}=-\frac{1}{2}+\frac{1}{2}\sqrt{1+f}\,,
\end{equation}
with $f$ depending on the quantum numbers as
\begin{equation} \label{eq:fconfdimorigin}
	f=4\,j^{+}\big(j^{+}+1\big)+\sum_{a} (2\pi\,p_{a})^{2}\,.
\end{equation}
The unitary bound $\Delta_{\rm L,R}=j^+$ is saturated for $p_{a}=0$ and the multiplets get shortened (see~\cite{Eloy:2021fhc}).

Turning on generic $(\omega,\zeta)$ deformations, the spectrum organises itself in representations of \eqref{eq: symmgeneric}.
The spectrum on arbitrary points of the family can be obtained by shifting the dimension of each physical mode in \eqref{eq:spectrumSU211} as $f\to f+\Delta f$ with
\begin{equation} \label{eq:confdimshift}
 	\Delta f=\frac{e^{2\omega}}{4}\Big((q_{\rm L}-q_{\rm R})+(q_{\rm L}+q_{\rm R})\,(e^{-2\omega}+\zeta^{2})+4\pi\,p_{7}\,\zeta\Big)^{2}-q_{\rm L}^{2}\,,
\end{equation}
where $q_{\rm L,R}$ denote the (integer-normalised) charges under ${\rm U}(1)_{\rm L,R}$, respectively.

In the heterotic case, the ${\cal N}=(0,4)$ supergroup organising the spectrum at the scalar origin is
\begin{equation} \label{eq:supergroupheterotic}
	\begin{aligned}
		\big[{\rm SU}(2)_{\rm l}\ltimes {\rm SU}(2)_{\rm L}\big]\times\big[{\rm SU}(2)_{\rm r}\ltimes &\text{SU}(2\vert1,1)_{\rm R}\big]\\
		&\times{\rm U}(1)^{4}\times{\rm SO}(16).
	\end{aligned}
\end{equation}
The spectrum follows from eq.~\eqref{eq:spectrumSU211} as a truncation that only keeps those states with integer spin under ${\rm SU}(2)_{\rm l}$, and further supplemented at each level by 16 copies of the multiplet
\begin{equation}
	\Big(\big(\Delta_{\rm L},0,j^{+}\big)\otimes\big[\Delta_{\rm R},0,j^{+}\big]\Big)_{p_{4},p_{5},p_{6},p_{7}},
\end{equation}
with $\Delta_{\rm L}=\frac{1}{2}+\frac{1}{2}\sqrt{1+f}$ and $\Delta_{\rm R}$ in \eqref{eq:confdimorigin}, forming an ${\rm SO}(16)$ vector. For the $(\omega,\zeta)$ deformation, the conformal dimension of each physical field gets shifted as in eq.~\eqref{eq:confdimshift}.

The masses $m_{(0)}$ of all scalars in the spectrum can be retrieved using eq.~\eqref{eq:spectrumSU211}--\eqref{eq:fconfdimorigin} and~\eqref{eq:confdimshift} on any point of the family of non-supersymmetric solutions \eqref{eq: twoparamin10D}, and analogously for the heterotic case. The Breitenlohner-Freedman bound~\cite{Breitenlohner:1982jf} in $3d$, \textit{i.e.} $\left(m_{(0)}\ell_{\rm AdS}\right)^2\geq-1$, shows that there are only two potentially unstable types of modes for each level $n=2\,j^{+}$, $p_{a}=0$ in \eqref{eq: Yfactorisation}. As conjectured in \cite{Eloy:2021fhc}, those modes have masses
\begin{equation}
	\begin{aligned}
	-(4+2\,n)+\left(2+n\right)^{2}\,e^{2\omega}\ [2]\,,\\
	 -(4+2\,n)+\left(2+n\right)^{2}\,e^{-2\omega}\,\left(1+e^{2\omega}\,\zeta^{2}\right)^{2}\ [2]\,,
	\end{aligned}
\end{equation}
with the integers between square brackets indicating their two-fold degeneracy. 
The stability condition is most requiring at the gauged supergravity level $n=0$. Thereby, the configuration is perturbatively stable if $(\omega,\zeta)$ lie within the range
\begin{equation} \label{eq: BFboundkkn0}
	e^{-\omega} \leq \dfrac{2}{\sqrt{3}}\,, \quad \
	\zeta^2 + \bigg(e^{-\omega}-\dfrac{\sqrt{3}}{4}\bigg)^{2} \geq \dfrac{3}{16}.
\end{equation}

In this Letter, Exceptional Field Theory has been utilised to find a family of non-supersymmetric deformations of the ${\rm AdS}_3\times S^3 \times {\rm T}^4$ solutions of heterotic and type II supergravities. We showed that these new solutions are perturbatively stable within a finite region of the parameter space and that there exists a one-dimensional subspace where ${\cal N}=(0,4)$ supersymmetry is preserved. Moreover, the holomorphicity arguments in the worldsheet formulation and boundary CFT$_2$ description suggest that these are solutions of string theory and not only purely large $N$ configurations. It will be interesting to explicitly compute the 1-loop corrections \`a la \cite{Bashmakov:2017rko,Fraiman:2023cpa} in the future to check this expectation.

The stability of the solution against non-perturbative decay channels needs to be investigated.
One possible decay channel for non-supersymmetric AdS$_3$ solutions is the destabilisation of the stack of branes that comprise them \cite{Ooguri:2016pdq,Apruzzi:2019ecr,Bena:2020xxb,Apruzzi:2021nle}.
We have explicitly checked that that is not the case for our non-supersymmetric solutions by considering probe F1- as well as D$p$- and NS5-branes with no worldvolume fluxes. These branes can be embedded in AdS$_3$, possibly wrapping the internal geometry, and their worldvolume actions show that they are attracted to the original stack, instead of emitted from it. 

Another possible decay channel is the nucleation of bubbles, including Coleman-de Luccia bubbles \cite{Coleman:1980aw} and bubbles of nothing \cite{Witten:1981gj,GarciaEtxebarria:2020xsr}. We leave this question for future work, but in line with the arguments in \cite{Giambrone:2021wsm}, one could expect the family to be protected due to the fact that it is away from the SUSY vacua only by a marginal deformation. 

\vspace{10pt}
\begin{acknowledgements}
We are grateful to Iosif Bena, Yolanda Lozano, Niall Macpherson, Emil Martinec, Chris N. Pope and Ergin Sezgin for helpful discussions and correspondence. 
We would also like to specially thank Emanuel Malek and Henning Samtleben for their feedback on a first version of this manuscript and collaboration on related projects. 
GL wants to thank the organisers of the `Supergravity, Strings and Branes' workshop at Bogazici University, Turkey, for giving him the opportunity to present this work. CE is supported by the FWO-Vlaanderen through the project G006119N and by the Vrije Universiteit Brussel through the Strategic Research Program ``High-Energy Physics''. GL is supported by endowment funds from the Mitchell Family \mbox{Foundation.}
\end{acknowledgements}

\bibliography{references}

\onecolumngrid

\end{document}